\documentclass[twocolumn]{aastex62}

\usepackage{ulem}
\hypersetup{linkcolor=red,citecolor=blue,filecolor=cyan,urlcolor=magenta}
\usepackage{amsmath}
\usepackage{tensor}     
\usepackage{graphicx}   
\usepackage{float}
\usepackage{bm}         
\usepackage{xcolor}     
\usepackage{color}      
\usepackage[section]{placeins} 
\usepackage[utf8]{inputenc} 
\newcommand{\nc}{\newcommand*} 
\nc{\figurewidth}{3.2in}
\nc{\xbar}{\bar{x}}
\nc{\rhoeq}{\rho_{\mathrm{eq}}}
\nc{\zeq}{z_{\mathrm{eq}}}
\nc{\tla}{\tilde{\lambda}}
\nc{\dt}{\delta}
\nc{\Dt}{\Delta}
\nc{\vj}{\bm{j}}
\nc{\vl}{\bm{l}}
\nc{\hx}{\hat{x}}
\nc{\hy}{\hat{y}}
\nc{\bj}{\bm{j}}
\nc{\mJ}{\mathcal{J}}
\nc{\mP}{\mathcal{P}}
\nc{\Msun}{M_\odot}
\nc{\app}{\approx}
\nc{\av}[1]{\langle #1 \rangle}
\nc{\eq}[1]{Eq.~\eqref{#1}}
\nc{\al}{\alpha}
\nc{\Xstar}{X_{\ast}}
\nc{\seq}{\sigma_{\mathrm{eq}}}
\nc{\fpbh}{f_{\mathrm{pbh}}}
\nc{\vth}{\bm{\theta}}
\nc{\vla}{\bm{\lambda}}
\nc{\vd}{\bm{d}}
\nc{\Mmin}{M_{\mathrm{min}}}
\nc{\rmd}{\mathrm{d}}
\nc{\mmin}{{m_{\mathrm{min}}}}
\nc{\mmax}{{m_{\mathrm{max}}}}
\nc{\mR}{\mathcal{R}}
\nc{\tmR}{\tilde{\mathcal{R}}}
\nc{\s}{\sigma}
\nc{\ogw}{\Omega_{\mathrm{GW}}}
\nc{\addref}{[\textcolor{red}{add ref}] }
\nc{\Om}{\Omega}
\nc{\gpcyr}{\mathrm{Gpc}^{-3}\,\mathrm{yr}^{-1}}
\nc{\Eq}[1]{Eq.~\eqref{#1}}
\nc{\Fig}[1]{Figure~\ref{#1}}
\nc{\Table}[1]{Table~\ref{#1}}
\nc{\lvc}{LIGO/Virgo} 
\nc{\Sec}[1]{Sec.~\ref{#1}}
\nc{\eg}{\textit{e.g.~}}
\nc{\SNR}{\mathrm{SNR}}

\def\({\left(}
\def\){\right)}
\def\[{\left[}
\def\]{\right]}

\def\e{\begin{equation}}
\def\q{\end{equation}}
\def\m{\begin{eqnarray}}
\def\n{\end{eqnarray}}

\received{}
\revised{}
\accepted{}
\published{}
\begin{document}

\title{Impact of a Spinning Supermassive Black Hole on the Orbit and Gravitational Waves of a Nearby Compact Binary}
\author{Yun Fang}
\email{fangyun@itp.ac.cn}
\affiliation{CAS Key Laboratory of Theoretical Physics, 
Institute of Theoretical Physics, Chinese Academy of Sciences,
Beijing 100190, China}
\affiliation{School of Physical Sciences, 
University of Chinese Academy of Sciences, 
No. 19A Yuquan Road, Beijing 100049, China}
\author{Xian Chen}
\email{xian.chen@pku.edu.cn}
\affiliation{Astronomy Department, School of Physics, Peking University, Beijing 100871, China}
\affiliation{Kavli Institute for Astronomy and Astrophysics at Peking University, Beijing 100871, China}
\author{Qing-Guo Huang}
\email{huangqg@itp.ac.cn}
\affiliation{CAS Key Laboratory of Theoretical Physics, 
Institute of Theoretical Physics, Chinese Academy of Sciences,
Beijing 100190, China}
\affiliation{School of Physical Sciences, 
University of Chinese Academy of Sciences, 
No. 19A Yuquan Road, Beijing 100049, China}

\date{\today}
\begin{abstract}
Recent theoretical studies suggest that stellar-mass binary black holes (BBHs)
	would merge more efficiently due to the Kozai-Lidov mechanism if these
	binaries form in the vicinity of supermassive black holes (SMBHs).
	Since SMBHs are likely rotating rapidly, we continue our earlier study
	on the generalization of the Kozai-Lidov formalism to include the spin
	of the SMBH and study the evolution of a nearby BBH.  We find that the
	eccentricity and orbital inclination of the BBH is significantly
	affected, because the spin (i) forces the orbital plane of the
	center-of-mass of the BBH around the SMBH to precess (the
	Lense-Thirring effect) and (ii) imposes an additional gravitomagnetic
	force on the BBH.  As a result, the  merger time of the BBH could be
	significantly different.  We calculate the waveform from the BBH in one
	representative example and study its detectability by a milli-Hertz GW
	detector, such as the Laser Interferometer Space Antenna (LISA). We
	find that the signal is distinguishable from that in the case without
	spin.  Our results imply that the BBHs in the LISA band could
	potentially be used to probe the spin of the SMBHs in galaxy centers. 
\end{abstract}

\keywords{gravitational waves --  
methods: data analysis -- celestial mechanics -- galaxies: nuclei 
} 

\section{Introduction}

The Laser Interferometer Gravitational-wave Observatory (LIGO) and the Virgo
detectors have detected ten binary black hole (BBH) mergers and one binary
neutron star inspiraling event during the first two observing runs
\citep{Abbott:2016blz, Abbott:2016nmj, Abbott:2017vtc, Abbott:2017gyy,
Abbott:2017oio, TheLIGOScientific:2016pea, TheLIGOScientific:2017qsa,
LIGOScientific:2018mvr}.  The origin of the LIGO/Virgo BBHs is unclear.  In the
conventional picture, BBHs form either in massive binary stars or star clusters
\citep{TheLIGOScientific:2016htt}.  Recent studies, however, suggest that the
centers of galaxies \citep{Miller:2008yw}, especially those hosting
supermassive black holes (SMBHs), are also important places for BBH formation
\citep[e.g.][]{Antonini:2012ad}.  In these environments, the merger rate of
BBHs could be enhanced \citep[see e.g.][and references therein]{Chen:2017xbi}.
The causes of this enhancement include a high escape velocity for compact
objects, a large concentration of black holes (BHs) due to the ``mass
segregation effect'', a tidal perturbation of the BBHs by the SMBHs, and a
hydrodynamical friction on each BH if gas is present.  The resulting merger
rate could be a significant fraction of the LIGO/Virgo event rate according to
recent calculations \citep{Hong:2015asd, VanLandingham:2016ccd, Hoang:2017fvh,
Petrovich:2017otm, Sedda:2018znc, Fragione:2018yrb, Bartos:2016dgn,
Stone:2016wzz, McKernan:2017umu}. 

According to these previous theoretical studies, a small fraction of BBHs
could either form at \citep{Inayoshi:2017hgw, Stone:2016wzz,
Bartos:2016dgn, McKernan:2017umu, Secunda:2018kar} or be captured to places
very close to the SMBHs \citep[e.g.][]{Addison:2015bpa, Chen:2018axp}.  As a
result, a triple system, composed of a BBH (as the ``inner binary'') revolving
around an SMBH (as the ``outer binary''), could form.  Because of the
perturbation by the SMBH, the inner BBH would undergo a ``Kozai-Lidov
oscillation'', during which the internal eccentricity of the BBH can be excited
to a large value \citep{Kozai:1962zz, Lidov:1962zs, Naoz:2016vh}. The
consequence is    a faster merger of the BBH \citep{Antonini:2012ad} or an
early detection of the BBH by a space-borne detector (such as the Laser
Interferometer Space Antenna, LISA) when the semi-major axis of the binary is
still large \citep{meiron_kocsis_2016,Hoang:2019kye, Randall:2019sab}. 

General relativistic effects are important during the evolution of the triple.
It has been shown that relativistic precession could suppress the Kozai-Lidov
(K-L) evolution, and that gravitational wave (GW) radiation could circularize
and limit the maximum eccentricity of the inner binary \citep{Wen:2002km,
Antonini:2012ad, Seto:2013wwa, VanLandingham:2016ccd, Fragione:2018nnl,
Chen:2018axp, Hoang:2019kye, Zhang:2019puc} .  However, these previous works
normally assume a Schwarzschild metric for the central SMBH. In reality, SMBHs
are spinning \citep[e.g.][]{Reynolds:2013qqa, Reynolds:2013rva,
Akiyama:2019cqa, Akiyama:2019bqs}.  The spin would induce a ``gravitomagnetic
field" \citep{Nichols:2011pu} in the spacetime, which causes an additional
precession to the outer orbit known as the ``Lense-Thirring effect''.  This
precession will invalidate the standard assumption in the K-L formalism that
the angular momentum of the outer orbit is effectively unchanged
\citep{Will:2017vjc}.  Moreover, the gravitomagnetic force also affects the
inner binary orbit, which is not included in the K-L formalism either.

Recently, we have extended the K-L formalism to include the spin effects and
found a modulation of the K-L cycle on a relatively long timescale
\citep{Fang:2019hir}.  Here we apply our method to study in more detail the
evolution of a stellar-mass BBH around a rotating SMBH.  We pay special
attention to the GWs emitted by the inner binary and look for imprint of the
spin.

The paper is organized as follows.  In Section~\ref{sec:evo}, we calculate the
dynamical evolution of a BBH around a spinning SMBH using our extended K-L
formulae including post-Newtonian (PN) corrections. More specifically, we
include pericenter precession (1PN), radiation reaction (2.5PN), and the most
importantly, the effects of spin acting on the orbits (1.5PN). We refer to this
model as ``K-L+1PN+RR+Spin'' and compare the results with that from the model
without spin, which we denote as ``K-L+1PN+RR''.  In
Section~\ref{sec:waveform}, as an example, we calculate the waveform of a BBH
around a SMBH similar to that in our Galactic Center.  We show in
Section~\ref{sec:FF} that in principle the results with and without spin are
distinguishable by LISA.  Finally, we summarize our findings in
Section~\ref{sec:con}.

\section{The dynamical evolution}\label{sec:evo}

\begin{figure}
\centering 
\includegraphics[width=0.4\textwidth]{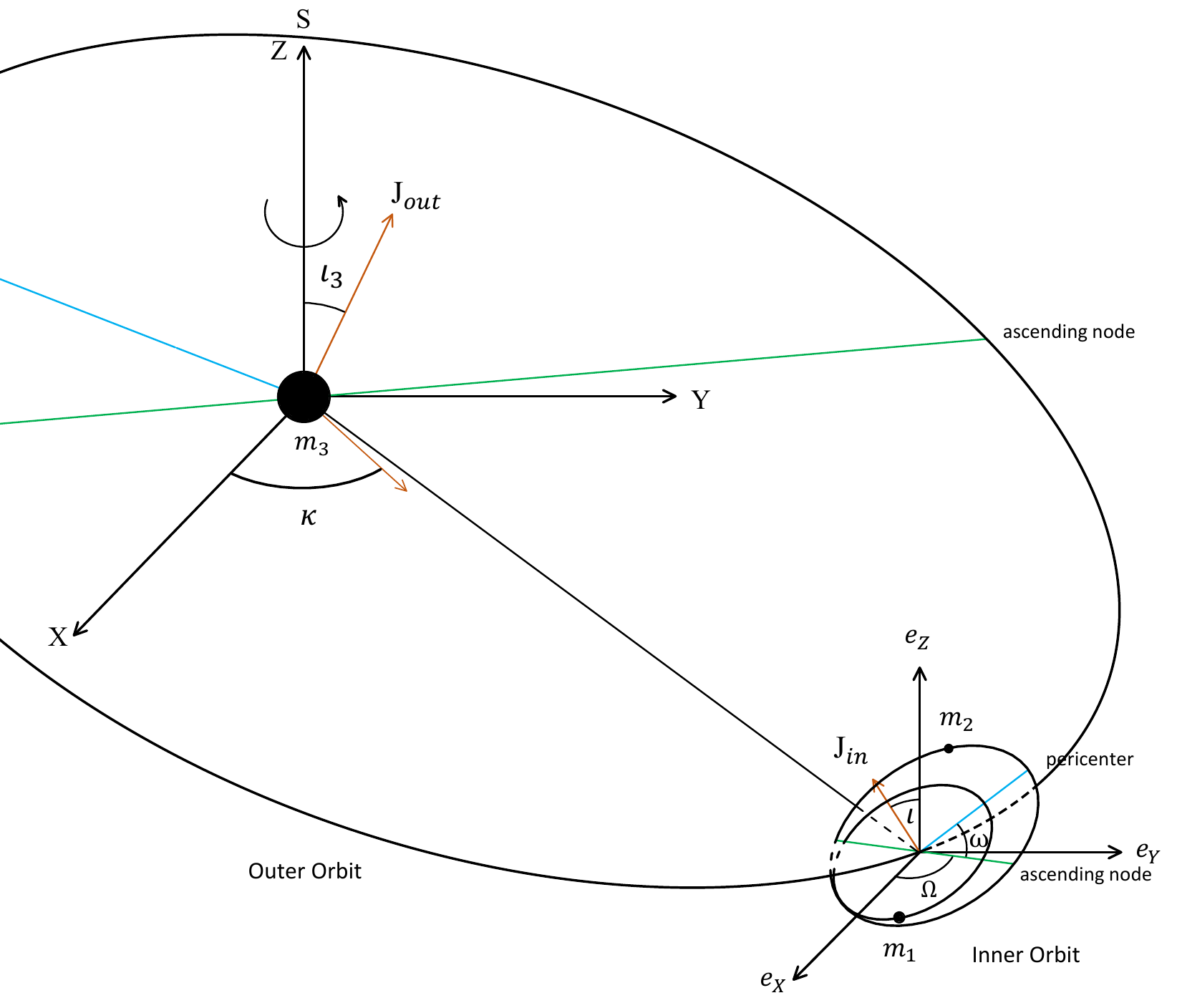}  
\caption{Configuration of our triple system seen from a fixed reference frame of $(X,Y,Z)$ with basis 
vectors $({\bf{e}}_X, {\bf{e}}_Y, {\bf{e}}_Z)$.
The vectors $\mathbf{J}_{\textnormal{out}}$ and $\mathbf{J}_{\textnormal{in}}$ are, respectively, 
the angular momenta of the outer and inner binary. Moreover,
$\iota_3$ and $\iota$ are the inclination angles of the outer and inner orbits, 
$\kappa$ is the angle between the projection of $\mathbf{J}_{\textnormal{out}}$ in the $(X,Y)$ plane 
and ${\bf{e}}_X$, 
$\Omega$ is the longitude of the ascending node of the inner orbit, and
$\omega$ the  pericenter angle. 
} 
\label{Figure.orbits} 
\end{figure}

The configuration of our triple system is illustrated in
Figure~\ref{Figure.orbits}.  Here, $m_1, m_2$ are the masses of the two BHs of the inner
binary, which is revolving around a SMBH with mass $m_3$.  The orbital
parameters are defined in a fixed coordinate system of $(X, Y, Z)$, with the
$Z$-axis aligned with the spin of the SMBH. Moreover,
$\mathbf{J}_{\textnormal{out}}$ and $\mathbf{J}_{\textnormal{in}}$ are,
respectively, the angular momenta of the outer and inner binaries.  The main
consequence of a rotating SMBH is to induce a gravitomagnetic field in which a
moving particle will feel a Lorentz-like force perpendicular to its velocity
\citep[e.g.][]{Thorne:1984mz,Nichols:2011pu,Poisson:2014aa}.  This
gravitomagnetic force affects both the outer and inner orbits
\citep{Fang:2019hir}. For the outer orbit, the net effect is a Lense-Thirring
precession of its angular momentum, which changes the angle $\kappa$ or,
equivalently, changes the longitude of the ascending node.  For the inner
orbit, the gravitomagnetic force modulates the inclination ($\iota$), ascending
node ($\Omega$), and pericenter ($\omega$), and in this way alters the K-L
oscillation. These effects are not present in the K-L formalism based on
non-spinning SMBHs. 

To illustrate the effects of the spin on the K-L oscillation, we show two
representative cases.  In both examples, we set $m_1=20\Msun$, $m_2=10\Msun$,
and $m_3=4\times10^6\Msun$, presenting the SMBH population similar to the one
in the Galactic Center. We also choose a spin value of $a=0.9m_3$ for the
central SMBH, where we have assumed $G=c=1$. The semi-major axis of the BBH is
chosen to be $\alpha=0.031$ AU, and that for the outer orbit is
$\mathcal{A}=30$ AU.  The corresponding eccentricities are, respectively, $
e=0.1$ and $E=0.1$. We choose these parameters so that the BBH and the SMBH
form a stable triple system, meanwhile the BBH is close
enough to the SMBH to be affected by its spin.

\begin{figure}
\centering 
\includegraphics[width=0.45\textwidth]{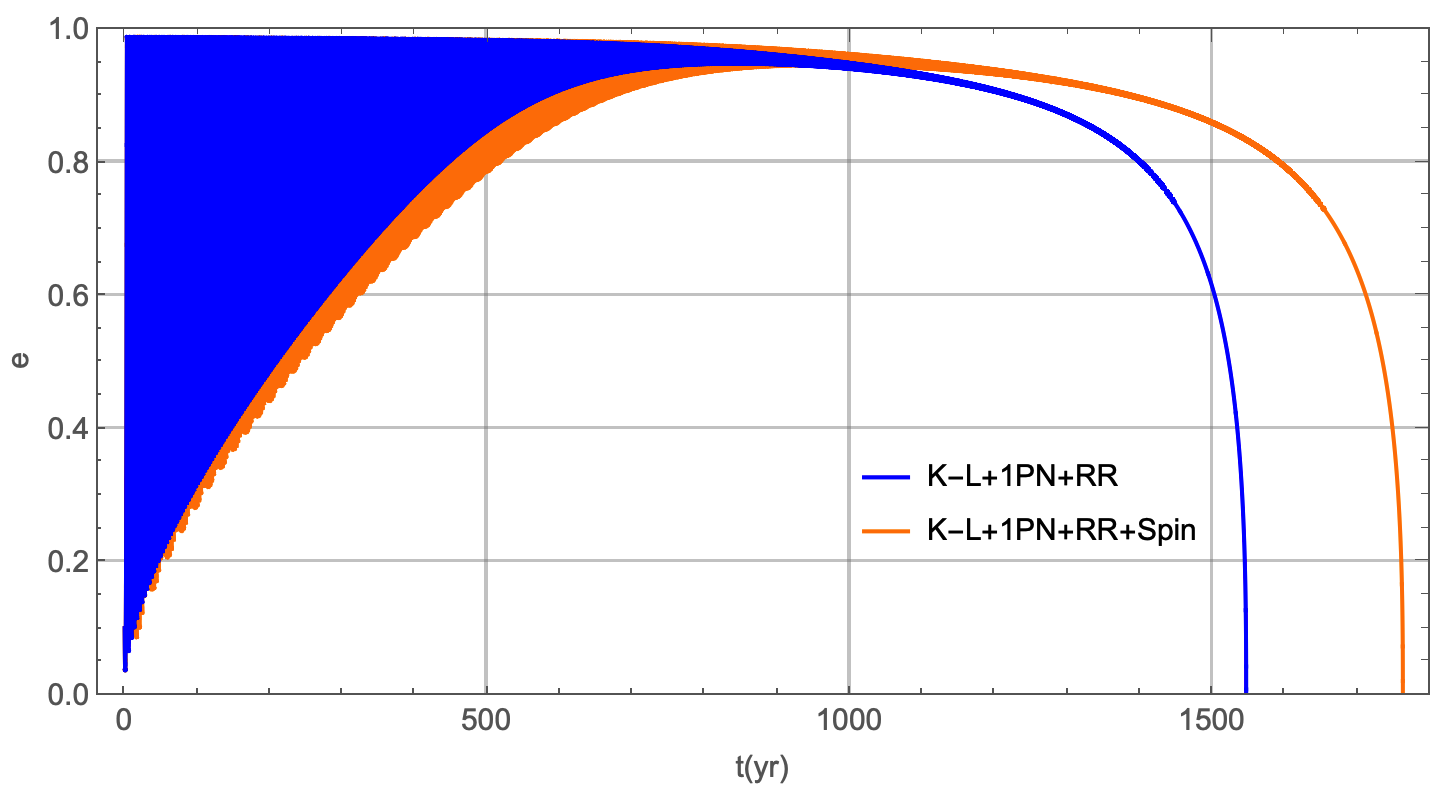}  
\caption{Evolution of the eccentricity of the BBH in our first example.
The orange curve is computed using our model including the spin effects, while the blue
one is computed without spin. } 
\label{Figure.eccentricityevolution} 
\end{figure}

In the first example, we use $\iota_3=60^{\circ}, \iota=70^{\circ},
\omega=20^{\circ}, \Omega=100^{\circ}$, and $\Omega_3=0$ as the initial
conditions, where $\Omega_3$ is the longitude of ascending node of the outer
orbit.  The resulting K-L oscillation is shown in
Figure~\ref{Figure.eccentricityevolution}. The orange curve refers to the
``K-L+1PN+RR+Spin'' model where the spin effects are included, and the blue one
refers to the ``K-L+1PN+RR'' model where the SMBH has zero spin.  We can see
that the eccentricity oscillates between $e=0.1$ and $e\simeq1$ at the
beginning of the evolution. This is a result of the K-L mechanism.  The
oscillation amplitude decreases with time because the suppression of the K-L
cycle by the 1PN precession becomes stronger as the semi-major axis decreases
due to GW radiation.  After about $10^3$ years the eccentricity no longer
oscillates because GW radiation starts to dominate the evolution of the BBH.
In this example, the merger time of the BBH is longer when spin is included. 

\begin{figure}
\centering 
\includegraphics[width=0.45\textwidth]{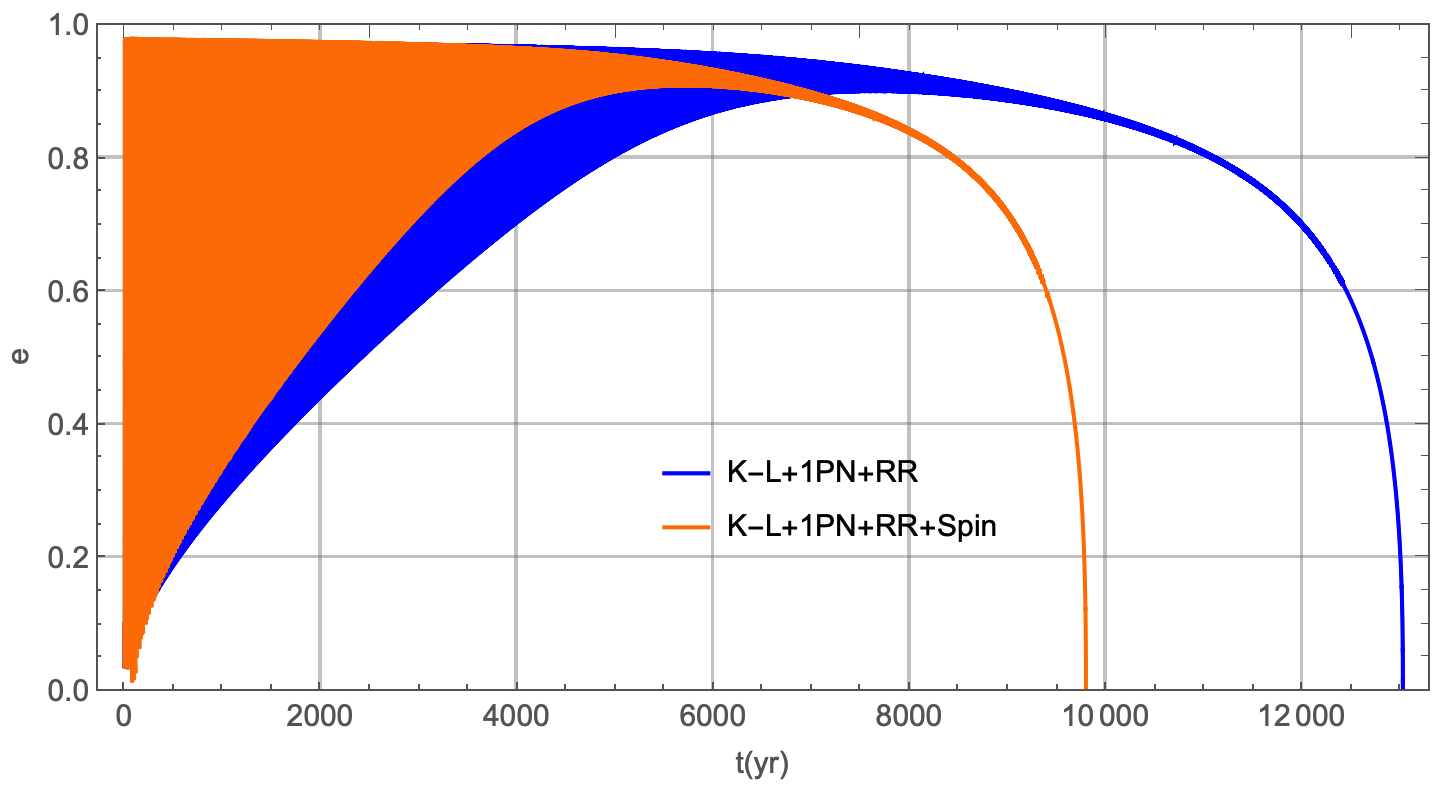} 
\caption{The same as Figure~\ref{Figure.eccentricityevolution} but for different angles in the
initial conditions.
} 
\label{Figure.eccentricityevolution2th} 
\end{figure}

In the second example, we change the initial angles to $ \iota=140^{\circ},
\omega=10^{\circ}, \Omega=60^{\circ}$, and keep the other parameters the same.
The results are shown in Figure~\ref{Figure.eccentricityevolution2th}.  In this
case, the presence of the spin shortens the lifetime of the BBH.

From the above two examples, we find that the spin affects the lifetime of the
BBH significantly, by about $14\%$ to $25\%$ relative to the lifetime around a
non-spinning SMBH. This difference could affect the event rate of BBH mergers
around SMBHs. We plan to study the impact on the LIGO/Virgo/LISA observations
in a future work.

\begin{figure}
\centering 
\includegraphics[width=0.45\textwidth]{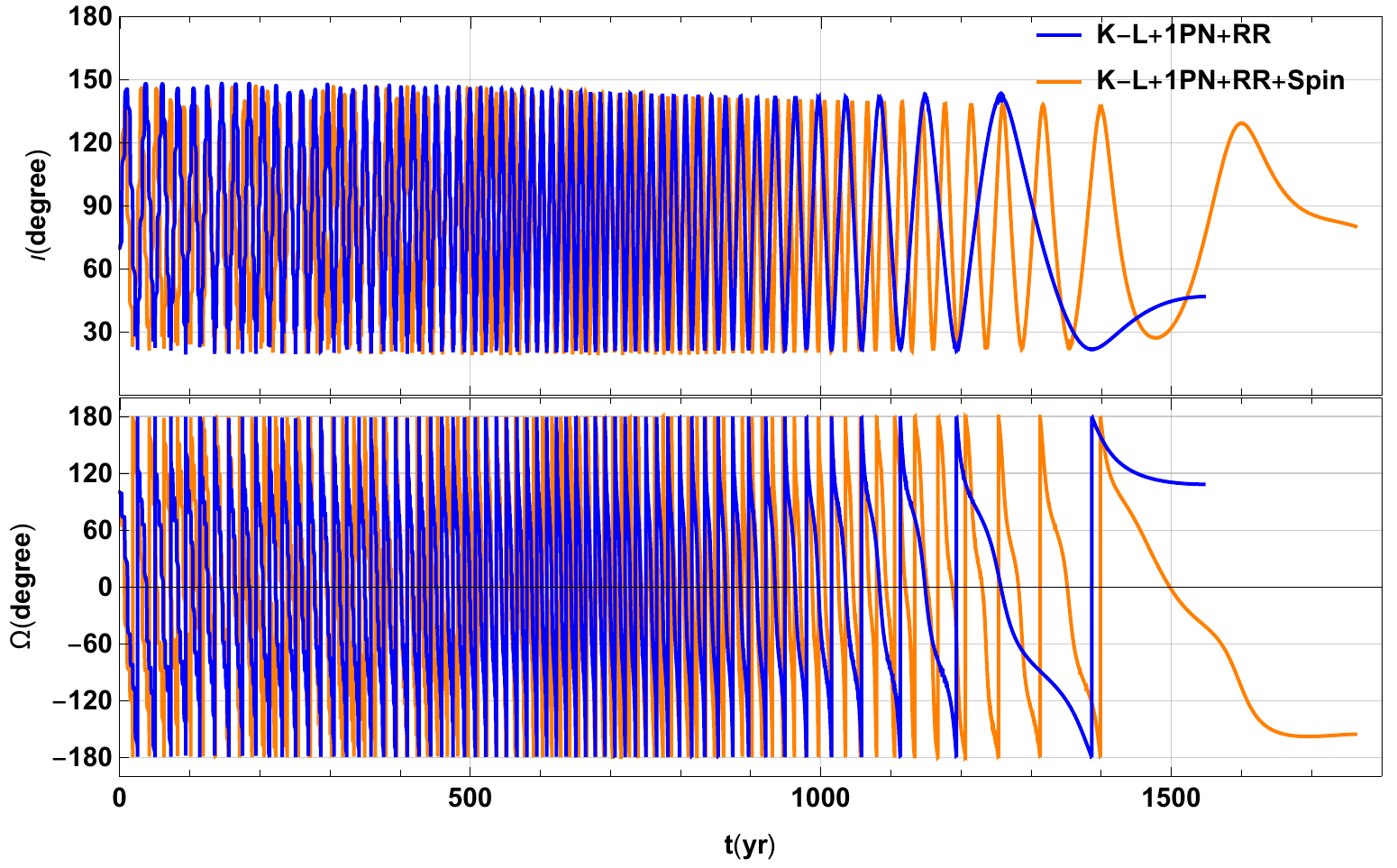}  
\caption{Evolution of $\iota$ and $ \Omega$. The line styles are the same as in Figure~\ref{Figure.eccentricityevolution}.  
} 
\label{Figure.iota_Omega_evolution} 
\end{figure}

\begin{figure}
\centering 
\includegraphics[width=0.45\textwidth]{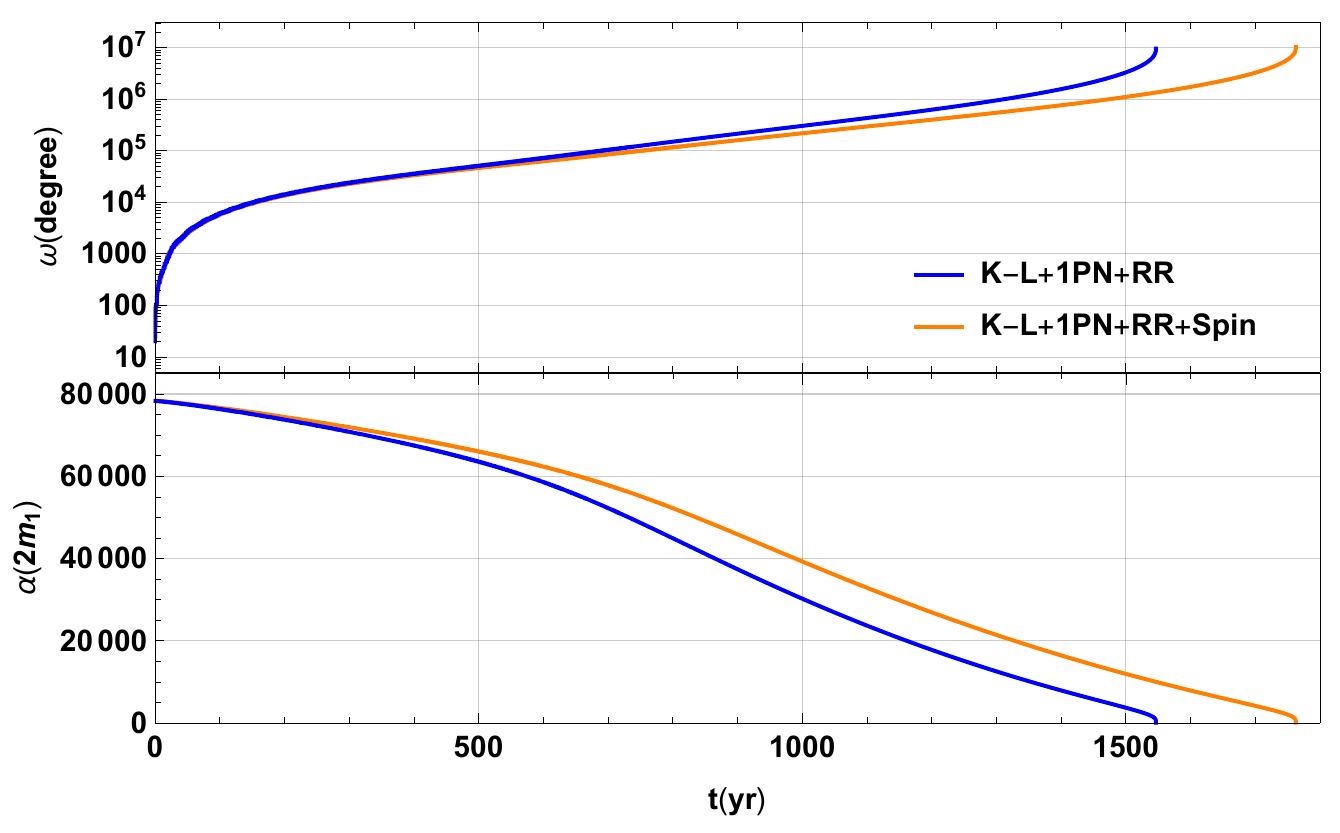}  
\caption{Evolution of $\omega$ and $ \alpha$, and the line styles are the same as in Figure~\ref{Figure.eccentricityevolution}. 
} 
\label{Figure.omega_alpha_evolution} 
\end{figure}

Here, we are interested in calculating the GWs generated by the inner binary.
Since the waveform is closely related to the projection of the two stellar BHs
in the plane perpendicular to the line-of-sight, we proceed to study the
orientation of the orbital plane (depending on $\iota$ and $\Omega$) and the
direction of the pericenter (depending on $\iota$, $\Omega$, and $\omega$) of
the inner orbit.  The results corresponding to our first example are shown in
Figures~\ref{Figure.iota_Omega_evolution} and
\ref{Figure.omega_alpha_evolution}.  We can see that $\iota$ oscillates more
frequently and $\Omega$ precesses faster when spin is added. By the end of our
simulation, the BBH coalesces with completely different $\iota$ and $\Omega$
compared to the case without spin. As for $\omega$, the precession is slower
when spin is included. The cause is that in the first example the semi-major
axis $\alpha$ shrinks more slowly when spin is present.

\begin{figure}
\centering 
\includegraphics[width=0.45\textwidth]{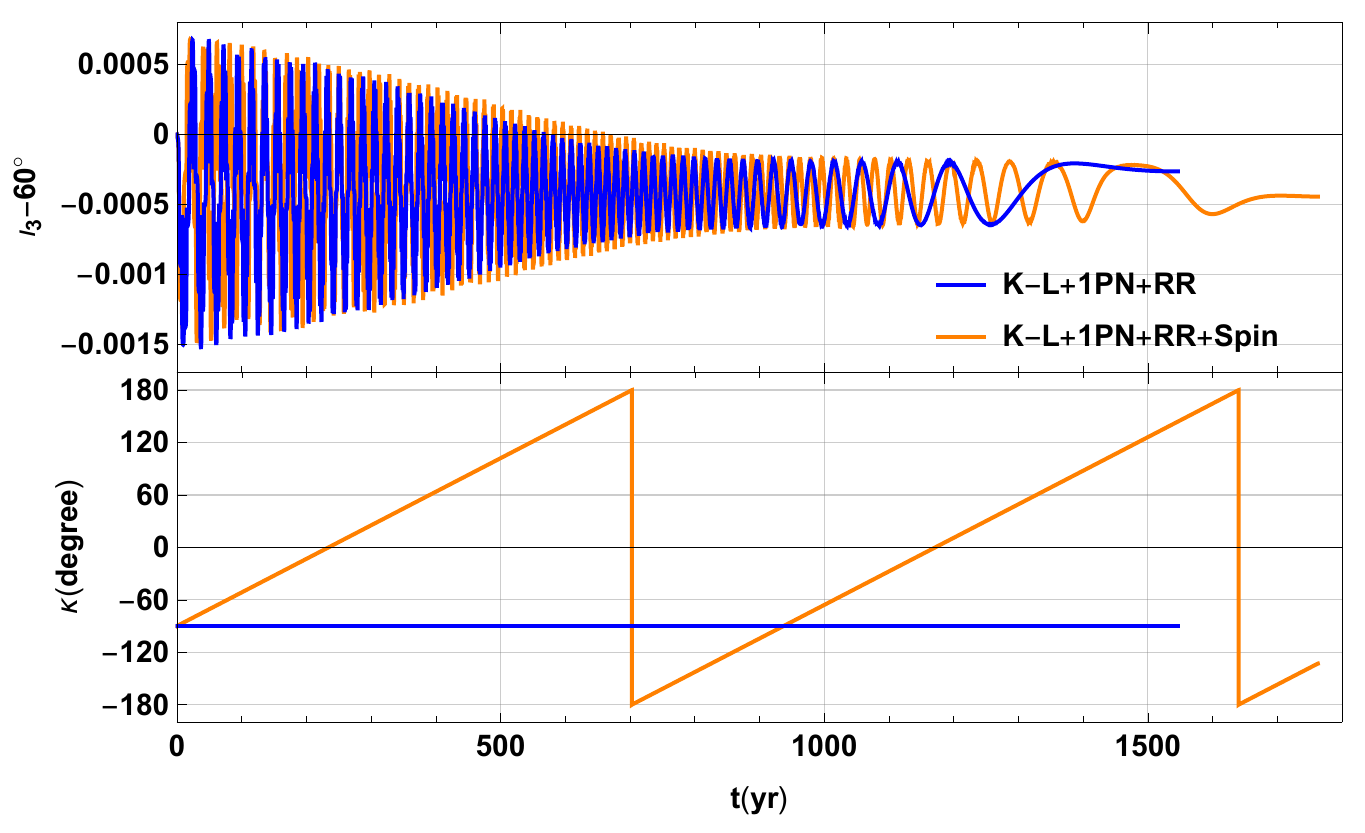}  
\caption{The same as  Figure~\ref{Figure.iota_Omega_evolution} but for $\kappa$  and $\iota_3$.  
} 
\label{Figure.iota3_kapa_evolution} 
\end{figure}

The spin of the SMBH also affects the outer orbit by inducing a Lense-Thirring
precession.  As can be seen in the lower panel of
Figure~\ref{Figure.iota3_kapa_evolution}, the consequence is a precession of
the angular momentum of the outer binary about the axis of the spin. We note
that the longitude of the ascending node (the aforementioned $\Omega_3$) is
equal to $\kappa+\pi/2$, and hence we do not plot it here. In the classic
picture of the Lense-Thirring effect, the inclination of the orbit ($\iota_3$)
is a constant. However, the upper panel of
Figure~\ref{Figure.iota3_kapa_evolution} shows that $\iota_3$ oscillates,
although the amplitude is small. The oscillation is caused by the coupling of
the inner and outer orbits in the K-L mechanism.  

The precession of the $\Omega_3$ angle of the outer orbit, in turn, affects the
dynamical evolution of the inner one. This is because $(\Omega-\Omega_3)$ and
$\iota_3$ enter the equations of motion of the inner binary
\citep{Fang:2019hir}.  In addition, the spin has a direct impact on  the
evolution of $\iota$, $\omega$, and $\Omega$. The combined effect causes a
different evolution of the inner binary in our “K- L+1PN+RR+Spin” model.

Moreover, the longitude of pericenter of the outer orbit also precesses due to
the spin.  However, it is decoupled from the evolution of the other orbital
elements, in the sense that it does not enter the equations of motion of the other orbital elements
 \citep[see e.g.][]{Naoz:2016vh, Will:2017vjc, Fang:2019hir},
and hence we could ignore it here.  Only from the octuple order does this angle
couple with the other orbital elements \citep[see e.g.][]{10.1093/mnras/stt302,
Naoz:2016vh}, but the coupling is small in our example because the ratio of the
two semi-major axis, $\alpha/\mathcal{A}$, is small.  

\section{Calculation of the waveform}\label{sec:waveform}

Now we use the angles derived in the last section to calculate the GW waveform
from the inner binary.  To facilitate the calculation, we assume that the
source is at a distance of $r$ from the detector,  the sky location is
$(\theta,\varphi)$, and the polarization angle is $\psi$ \citep[see][ for the
definition of the angles]{Apostolatos:1994mx}.  We further denote
$(\gamma,\beta)$ as the angles describing the orientation of the wave vector in
the source frame $(x, y, z)$, which is defined in such a way that the $(x,\,y)$
plane aligns with the orbital plane of the BBH, the $x$-axis is in the
direction of the pericenter, and the $z$-axis aligns with
$\mathbf{J}_{\textnormal{in}}$.  In this way, the $(x, y, z)$ frame is linked
to the $(X, Y, Z)$ coordinate system by a Euler transformation with the angles
$(\omega, \iota, \Omega)$.  To calculate $(\gamma,\beta)$, we fix the
line-of-sight along the $X$-axis, for simplicity.  As a result, we have
\begin{align}
\cos \gamma&={\bf{e}}_z \cdot {\mathbf{e}_X}=\sin \iota  \sin \Omega, \\
\sin \gamma&={({{\mathbf{e}}_X-({\mathbf{e}}_z \cdot {\mathbf{e}_X}) {\mathbf{e}_z}})\cdot {\mathbf{e}_X}\over |{{\mathbf{e}}_X-({\mathbf{e}}_z \cdot {\mathbf{e}_X}) {\mathbf{e}_z}}|} \nonumber\\
&=\frac{2-2 \sin ^2\iota\sin ^2\Omega}{\sqrt{2 \cos 2 \iota  \sin ^2\Omega+\cos 2 \Omega +3}}, 
\end{align}
and 
\begin{align}
\cos \beta&=\frac{2 \cos \omega  \cos \Omega -2 \cos \iota  \sin \omega \sin \Omega }{\sqrt{2 \cos 2\iota  \sin ^2\Omega +\cos 2 \Omega +3}}, \\
	\sin \beta&=-{2 \cos \iota \cos \omega \sin \Omega+2 \sin \omega \cos \Omega \over \sqrt{2 \cos 2 \iota  \sin ^2\Omega+\cos 2 \Omega+3}}\label{eq:sign}
\end{align}
\citep[see also Eq.~(11) in][]{Fang:2019hir}.

The strains of the $+$ and $\times$ polarizations of the GWs can now be
calculated with
\begin{align}
	h_{+}&={1\over 2}(\bm{\gamma}^j \bm{\gamma}^k - \bm{\beta}^j \bm{\beta}^k) h^{\textnormal{TT}}_{jk}, 
	\label{h_plus}
	\end{align}
\begin{align}
	h_{\times}&={1\over 2}(\bm{\gamma}^j \bm{\beta}^k + \bm{\beta}^j \bm{\gamma}^k)h^{\textnormal{TT}}_{jk},
	\label{h_cross}
\end{align}
where $\bm{\gamma}:=(\cos\gamma \cos\beta, \cos\gamma \sin\beta, -\sin\gamma)$
and $\bm{\beta}:=(-\sin\beta,\cos\beta,0) $. Here, $h^{\textnormal{TT}}_{jk}$
is the $\textnormal{TT}$ projection of the GW strain $h_{jk}$.  The GW
waveform of an eccentric orbit is composed of many Fourier modes \citep[see
e.g.][]{Maggiore2007}, and the $n$-th mode can be calculated with 
\begin{eqnarray}
	    h_{+, n}&=& -\frac{\pi ^2 {f}^2 \mu  n^2 }{r}\{ 2 {C_{n}} \sin 2 {\beta} (\cos 2{\gamma}+3) \sin (2 \pi  n f t) \nonumber\\
&+&\cos (2 \pi  n f t) [({A_{n}}-{B_{n}}) \cos 2{\beta} (\cos 2 {\gamma}+3)  \nonumber\\
&-&2 ({A_{n}}+{B_{n}}) \sin^2{\gamma} ] \}, \\
	    h_{\times, n}&=& -\frac{4 \pi ^2 {f}^2 \mu  n^2 \cos {\gamma}}{r} [({B_{n}}-{A_{n}}) \sin 2{\beta} \cos (2 \pi  n f t) \nonumber\\
&+&2 {C_{n}} \cos 2 {\beta} \sin (2\pi  n f t)], \\
\end{eqnarray}
where $f$ is the radial orbital frequency of the inner binary, $\mu=m_1 m_2/(m_1+m_2)$ is the reduced mass,
and
\m
A_{n}&=&\frac{{\alpha}^{2}}{ n} [J_{n-2}\left(n e\right)-J_{n+2}\left(n e\right) -2 e J_{n-1}\left(n e\right) \nonumber\\
&+&2 e J_{n+1}\left(n e\right)] , \\
B_{n}&=&-\frac{\left(1-e^2\right) {\alpha}^{2}}{ n} [J_{n-2}\left(n e\right)-J_{n+2}\left(n e\right)] , \\
C_{n}&=&\frac{\sqrt{1-e^2} {\alpha}^{2}}{ n} [J_{n-2}\left(n e\right)+J_{n+2}\left(n e\right) \nonumber\\
&-&e J_{n-1}\left(n e\right)-e J_{n+1}\left(n e\right)], \\ 
\n

\begin{figure}
\centering 
\includegraphics[width=0.45\textwidth]{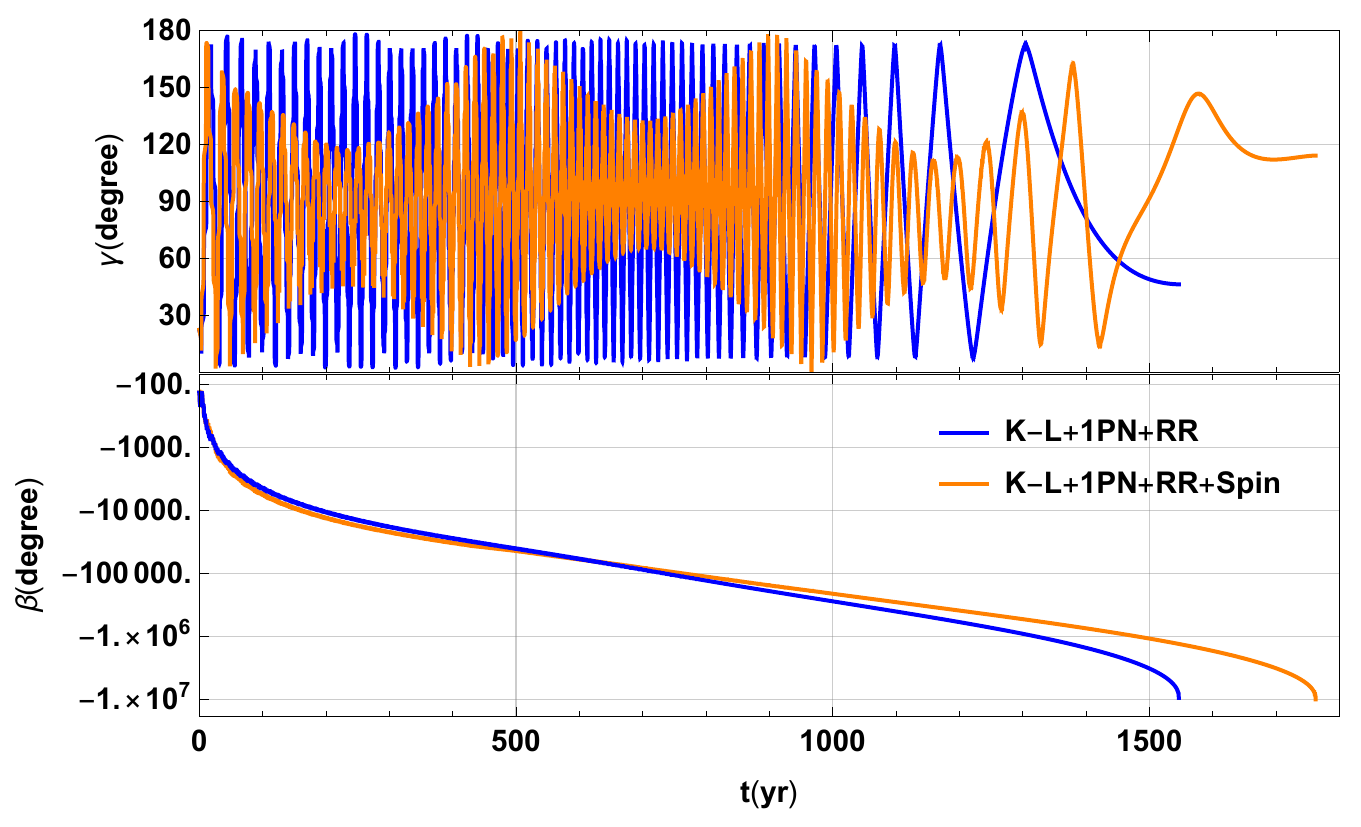}  
\caption{Evolution of $\gamma$ (upper panel) and  $\beta$  (lower panel). The blue curves 
	correspond to the model without spin and the orange ones with spin.} 
\label{Figure.gamma_beta_evolution} %
\end{figure}

We can see that the waveform is closely related to the angles $\gamma$ and
$\beta$.  Figure~\ref{Figure.gamma_beta_evolution} shows the difference of
$\gamma$ and $\beta$ in our first example.  We find that in this example,
$\gamma$ changes in two ways when spin is included.  First, it oscillates with
a higher frequency. Second, its amplitude shows a periodic modulation. The
first effect is caused by a combination of the Lense-Thirring effect on the
outer orbit and the gravitomagnetic force acting on the inner orbit.  The
latter effect, i.e., the periodic modulation of the amplitude, is closely
related to the Lense-Thirring precession of the outer orbit, since the
timescale is the same as the precession of $\kappa$ shown in
Figure~\ref{Figure.iota3_kapa_evolution}. The behavior of $\beta$ is similar to
$\omega$ in Figure~\ref{Figure.omega_alpha_evolution}. The difference is
mainly a minus symbol because in the source frame the direction of the
precession reverses.

We note that the revolution of the BBH around the SMBH, in principle, also
generates GWs \citep[e.g. see an example in][]{Chen:2018axp}.  We do not
calculate them because for the parameters considered in this work the
corresponding frequency is much lower than mHz, and hence outside the LISA
band. Moreover, the revolution should also periodically modulate the phase of
the GWs (from the BBH), as a consequence of the Doppler frequency shift
\citep{Inayoshi:2017hgw,meiron_kocsis_2016}, as well as modulate the amplitude
due to the Lorentz transformation of the wave vectors \citep{torres18}. We do
not include these effects in our calculations because in our examples they are
of PN order; they are secondary effects relative to the modulation of $\gamma$
and $\beta$.

\section{Matched filtering}\label{sec:FF}

\subsection{General consideration}\label{subsec:FF1}

\begin{figure}
\centering 
\includegraphics[width=0.45\textwidth]{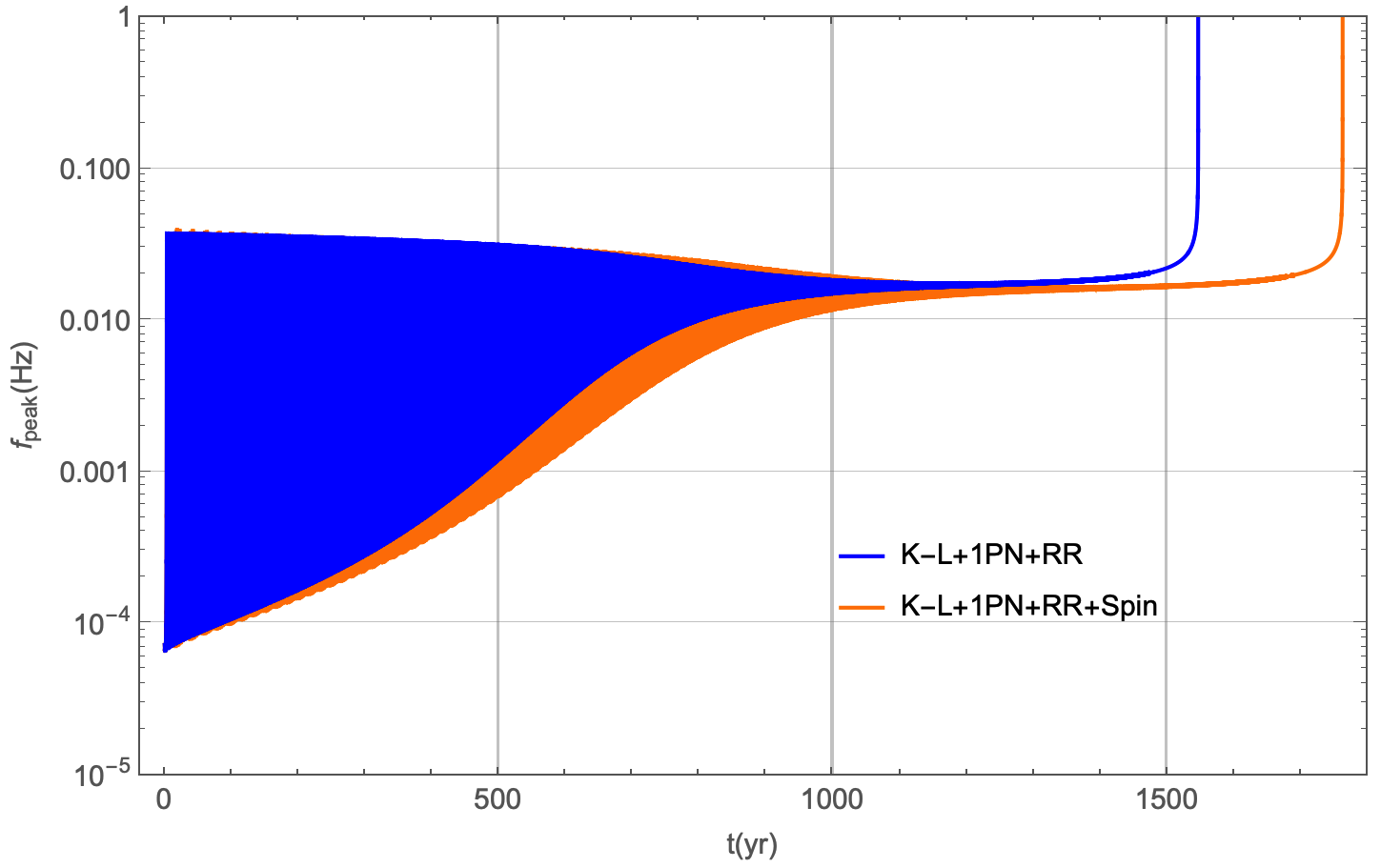}  
\caption{Evolution of the peak frequency of the GWs from the inner BBH. The line styles 
and the model parameters are the same as in Figure~\ref{Figure.eccentricityevolution}.
} 
\label{Figure.peakfrequency} 
\end{figure}

Now we study whether or not LISA could detect the spin of the SMBH from the
waveform of the BBH.  
We take the first example in Section~\ref{sec:evo} for illustrative 
purposes because, as we will show below, the BBH evolves into 
the LISA band on a relative short timescale.
It is known that a highly eccentric BBH emits a broad spectrum of GWs
\citep{Peters:1963ux}. The peak of this spectrum is located at a frequency
closely linked to the pericenter distance $\alpha(1-e)$, i.e.,
\e
f_{\rm peak}={\sqrt{m_1+m_2}(1+e)^{-0.3046} \over \pi \left[\alpha(1-e)\right]^{3/2}}
\q
\citep{Wen:2002km}.  If this peak enters the LISA band ($10^{-3}-10^{-1}$ Hz),
the BBH could be detected.  Figure~\ref{Figure.peakfrequency} shows the
evolution of $f_{\rm peak}$ in our first example.  We find that the BBH dwells
in the LISA band during the first $1500-1700$ years of the evolution.
Afterwards, the BBH moves into the LIGO/Virgo band and coalesces.

\begin{figure*}
\centering 
\includegraphics[width=0.85\textwidth]{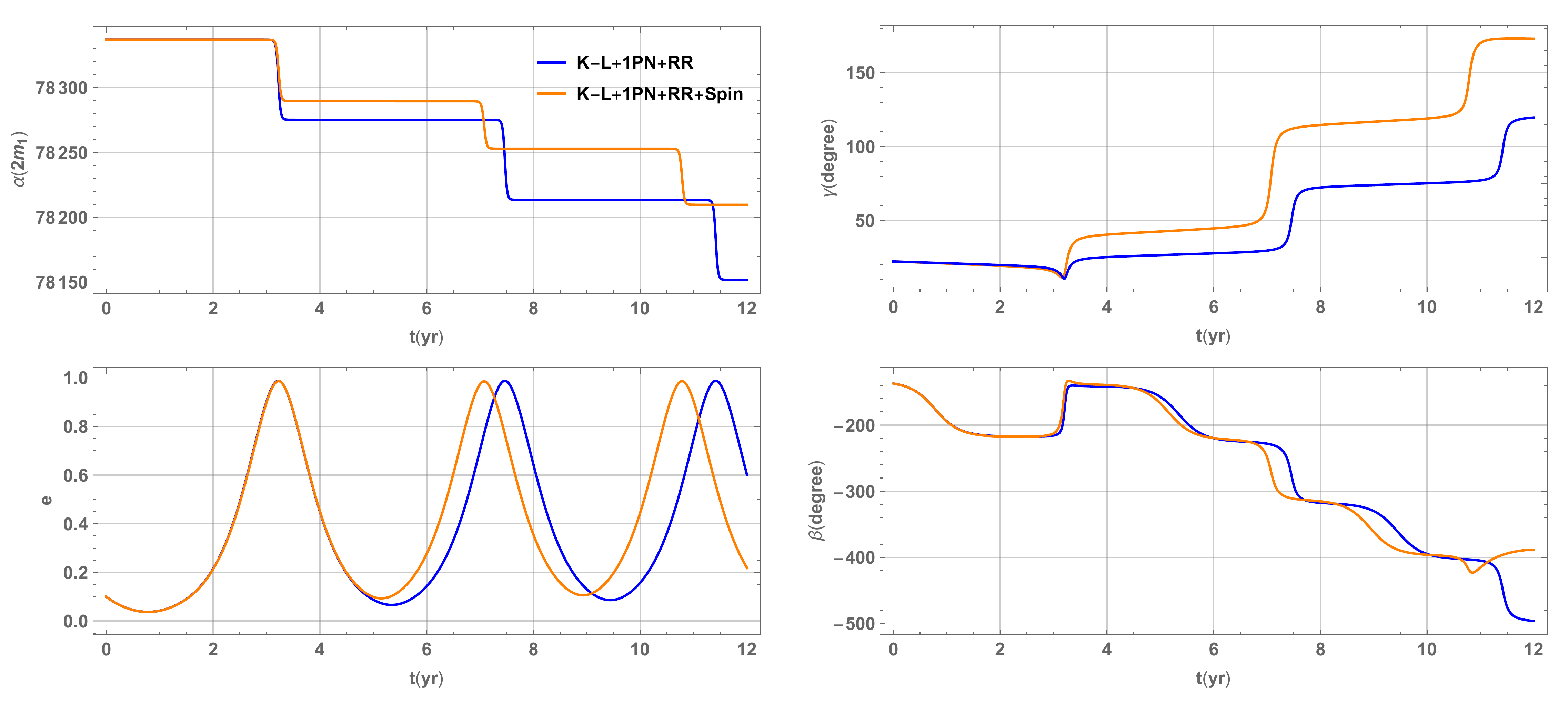}  
\caption{Evolution of $e, \alpha, \gamma, \beta$ in the first 12 years for our first example. 
	The unit of $\alpha$ is $2m_1$, i.e., the Schwarzschild radius of the more massive BH in the binary.  } 
\label{Figure.inner_orbit_12yr} 
\end{figure*}

To see more details of the K-L evolution during a period comparable to the
observational timescale, we shown in Figure~\ref{Figure.inner_orbit_12yr} the
evolution of  $\alpha$, $e$, $\gamma$, and $\beta$ during the first 12 years.
We find that the K-L timescale is about four years and it becomes shorter when
spin of the SMBH is introduced. This result indicates that (i) LISA could
detect the entire K-L cycle if the observational period is longer than four years and (ii)
the difference of the waveform induced by the spin of the SMBH could be
detectable by LISA.

In practice, LISA employs a method called the ``matched filtering'' to discern
the difference between two waveforms \citep{Finn:1992wt, Cutler:1994ys,
Lindblom:2008cm}. For example, given $h_1$ and $h_2$, the inner product is
defined as 
\e
\langle h_1| h_2 \rangle=2 \int_{0}^{\infty} {{\tilde{h}^*_1(f)\tilde{h}_2(f)+\tilde{h}_1(f)\tilde{h}^*_2(f)} \over S_{h}(f)} df,\label{eq:inner} 
\q
where
 $S_{h}(f)$ is the spectral noise density \citep[Eq.~(1) in][]{Cornish:2018dyw}  and $\tilde{h}_i (f)$ denotes the Fourier transformation
\e
\tilde{h} (f)=\int_{-\infty}^{\infty} e^{2\pi i f t}h (t)dt. 
\label{fouriertrans}
\q
The two waveforms are indistinguishable if the condition
\e
\langle \delta h|\delta h \rangle=\langle h_1-h_2| h_1-h_2 \rangle<1 
\label{waveformdistincondition}
\q
is satisfied.

In real observations, we are often in a situation where $h_1\simeq h_2$.  In
this case, the calculation of Equation~(\ref{waveformdistincondition}) can be
significantly simplified. One can, instead, calculate a quantity called the
``fitting factor'' (\textnormal{FF}) \citep{Apostolatos:1994mx,
Lindblom:2008cm}, defined as  
\e
\textnormal{FF}={\langle  h_1| h_2 \rangle \over \sqrt{ \langle  h_1| h_1 \rangle \langle  h_2| h_2 \rangle} }, 
\label{fittingfactor}
\q
and compare it with a threshold \textnormal{FF} defined as
\e
\textnormal{FFS}=1-{1\over \langle  h_1| h_1 \rangle+\langle  h_2| h_2 \rangle}.
\q
The latter \textnormal{FFS} is closely related to the signal-to-noise ratio
(SNR), i.e., ${\rm SNR}=\sqrt{\langle h| h \rangle}$.  If $h_1\simeq h_2$, we
have $\textnormal{FFS} \simeq1-1/(2\,{\rm SNR}^2)$, and the criterion of
Equation~(\ref{waveformdistincondition}) is equivalent to
$\textnormal{FF}>\textnormal{FFS}$.  For example, LISA will claim a detection
as soon as ${\rm SNR}$ becomes $10$.  This means for LISA sources,
\textnormal{FFS} is at least $0.995$, and only the waveform template with
$\textnormal{FF}>0.995$ is an acceptable match to the signal.

To account for high eccentricities, the inner product defined in
Equation~(\ref{eq:inner}) can be computed using the harmonics of $h_1$ and
$h_2$, i.e.,
\e
\langle h_1| h_2 \rangle=2\sum_{n=1}^{\infty} {{\tilde{h}^*_{1, n}(f_n)\tilde{h}_{2, n}(f_n)+\tilde{h}_{1, n}(f_n)\tilde{h}^*_{2, n}(f_n)} \over S_{h}(f_n)} d f_n. 
\q

We calculate the frequency of the $n$-th harmonic with the approximation
$f_n\simeq n f_0$, where $f_0$ is the radial orbital frequency. It is an
approximation because we have neglected the shift of the GW frequency caused by
the precession of the pericenter. The approximation is acceptable here because
the shift of $f_n$ is $\dot{\omega}/\pi$ \citep{Barack:2003fp}, which is of PN
order and is much smaller than $f_0$.

\subsection{Specific implementation}

In our own problem, the difference between $h_1$ and $h_2$ is caused by the
spin of the SMBH.  For the convenience of the following analysis, we denote the
waveform calculated from the ``K-L+1PN+RR" model as $h_1(e_1, \alpha_1,
\gamma_1, \beta_1)$  and that from the ``K-L+1PN+RR+Spin" as $h_2(e_2, \alpha_2
, \gamma_2, \beta_2)$. Each waveform is a weighted sum of the two polarizations
defined in Section~\ref{sec:waveform}, i.e.,
\e
\begin{split}
h (t)=&F_{+}(\theta,\varphi,\psi) {\sqrt{3}\over 2}h_{ +}(t, r, e, \gamma, \beta)\\
&+F_{\times}(\theta,\varphi,\psi) {\sqrt{3}\over 2} h_{ \times}(t,r, e, \gamma,\beta), 
\end{split}
\label{respondstrain}
\q
where $F_+$ and $F_{\times}$ are the ``antenna patterns''.  The coefficient
$\sqrt{3}/2$ comes from the fact that the actual angle between LISA arms is
$60^{\circ}$ \citep{Berti:2004bd, Nishizawa:2016jji}.  For simplicity,  we
average the source location angles ($\theta$ and $\varphi$) and the
polarization angle ($\psi$), so that we do not need to consider the motion of the
LISA arms \citep[also see][]{Barack:2003fp, Cornish:2018dyw}.  As a result of
the average, we have $\langle F_{+}^2\rangle=\langle F_{\times}^2\rangle=1/5$ and
$\langle F_{+}F_{\times}\rangle=0$ \citep{Flanagan:1997sx, Cornish:2018dyw}. 

We notice that in this example,  the radial orbital frequency of the BBH in the
first four years (the first K-L cycle) of evolution is almost a constant.  Therefore, for each
Fourier mode $S_h(f_n)$ is also a constant.  In this case, by Parseval's
theorem, the integration in the frequency domain can be performed in the time
domain \citep{Barack:2003fp},
\begin{align}
	&\langle h_1| h_2 \rangle=2\sum_{n=1}^{\infty} {1\over S_{h}(f_n)}\int_{0}^{\Delta t} {|{h}_{1, n}(t){h}_{2, n}(t)|} d t\\
	&= \sum_{n=1}^{\infty} {3/10 \over S_{h}(f_n)}\int_{0}^{\Delta t} (|{h}_{1+, n}
	{h}_{2+, n}|+|{h}_{1\times, n}{h}_{2\times, n}|) d t, 
\end{align}
where we have applied the result $\langle F_{+}F_{\times}\rangle=0$ derived
above so that the cross terms with ${h}_{1+, n}{h}_{2\times, n}$ and
${h}_{1\times, n}{h}_{2+, n}$ vanish. Moreover, the results for ${|h_{+,
n}|}^2+{|h_{\times,n}|}^2$ and
${|h_{1+,n}h_{2+,n}|}+{|h_{1\times,n}h_{2\times,n}|}$  do not depend on the
definition of the GW polarizations in Equation \ref{h_plus} and \ref{h_cross}
as long as we averaged out $F_+ and F_{\times}$ \citep{Poisson:2014aa}.  
By averaging ${|h_{+,n}|}^2+{|h_{\times,n}|}^2$ and
${|h_{1+,n}h_{2+,n}|}+{|h_{1\times,n}h_{2\times,n}|}$ over GW phase, 
we have \citep[following][]{Maggiore2007}
\begin{eqnarray} &&{|h_{1+,n}|}^2+{|h_{1\times,n}|}^2 = \frac{\pi ^4 f^4
	    \mu ^2 n^4}{2 r^2}  ( 4 {C_{1n}}^2 \sin ^2 2 \beta _1 (\cos 2
	    \gamma _1+3){}^2  \nonumber\\ &&+(({A_{1n}}-{B_{1n}}) \cos 2 \beta
	    _1 (\cos 2 \gamma _1+3)-2 ({A_{1n}}+{B_{1n}}) \sin ^2\gamma _1){}^2
	    \nonumber\\ &&+16 \cos ^2\gamma _1 (({A_{1n}}-{B_{1n}})^2\sin ^22
	    \beta _1+4 {C_{1n}}^2 \cos ^2 2 \beta _1) ), \\ \nonumber\\
	    &&{|h_{2+,n}|}^2+{|h_{2\times,n}|}^2 =
	    ({|h_{1+,n}|}^2+{|h_{1\times,n}|}^2 ) (1\to2),  \\ \nonumber\\
	    &&{|h_{1+,n}h_{2+,n}|}+{|h_{1\times,n}h_{2\times,n}|} = \nonumber\\
	    &&{\pi ^4 f^4 \mu ^2 n^4 \over 2 r^2}(16 \cos {\gamma _1} \cos
	    {\gamma _2} (4 \cos {2 \beta _1} \cos {2 \beta _2} C_{1n} C_{2 n}
	    \nonumber\\ &&+\sin {2 \beta _1} \sin {2 \beta _2} (A_{1n}-B_{1n})
	    (A_{2 n}-B_{2 n})) \nonumber\\ &&+( \cos {2 \beta _1} (\cos {2
	    \gamma_1}+3) (A_{1n}-B_{1n})-2 \sin ^2{\gamma _1} (A_{1n}+B_{1n}))
	    \nonumber\\ &&\times ( \cos {2 \beta _2} (\cos {2 \gamma _2}+3)
	    (A_{2 n}-B_{2 n})-2 \sin ^2{\gamma _2}(A_{2 n}+B_{2 n}))
	    \nonumber\\ &&+4 \sin {2 \beta _1} \sin {2 \beta_2} (\cos 2 \gamma
    _1+3) (\cos {2 \gamma_2}+3) C_{1n} C_{2 n}), 
	  \label{hh_equations}
\end{eqnarray}

With the above preparations, we can proceed to calculate the FF defined in
Equation~(\ref{fittingfactor}).  Assuming that our source is in the Galactic
Center, at a distance of $r=8~{\rm kpc}$, and LISA observes it for four years,
we find for our first example that $\textnormal{FF}=0.834$ and
$\textnormal{FFS}=0.999997$. The result $\textnormal{FF}<\textnormal{FFS}$
indicates that if there is such a BBH around the SMBH in the Galactic Center,
LISA should be able to detect the effects induced by the spin. 


\section{Conclusions}\label{sec:con}

In this paper we have studied the impact of the spin of a supermassive black
hole (SMBH) on the orbital evolution of a nearby compact binary.  Our model is
based on our previous theoretical work on the extension of the Kozai-Lidov
formalism to include the effects due to the spin of the tertiary body
\citep{Fang:2019hir}. By comparing the orbital evolution of the BBH in the
cases with and without the spin effects, we find the following results from our
representative examples.  (i) When the spin is present, the dynamical evolution
of the BBH is significantly different, mainly caused by the Lense-Thirring
precession of the outer orbit and the gravitomagnetic force acting on the inner
orbit. (ii) The merger time of the BBH could be elongated or shortened by the
presence of the spin,  depending on the initial orbital angles. (iii) The
combined effects of the Lense-Thirring precession (of the outer binary) and the
gravitomagnetic force (on the inner binary) causes the inclination ($\gamma$)
of the inner orbit relative to the line-of-sight to oscillate more rapidly,
while the Lense-Thirring precession also periodically modulates the amplitude
of the oscillation of $\gamma$ on the same timescale.

We have also developed an analytical framework to calculate the GWs from the
BBHs in our problem and used it to study the impact of the spin on LISA
waveforms. We find the following differences from our representaive example.
(i) The polarization angle (azimuth angle $\beta$) of the waveform precesses
differently when spin is included in our calculation.  (ii) The period of the
Kozai-Lidov oscillation, and hence the timescale on which the BBH enters and
exits the LISA band, changes as a result of the spin of the SMBH.  (iii) Given
the triple in our representative example, we find that
$\textnormal{FF}<\textnormal{FFS}$ when the SNR of the source is high enough,
which indicates that for such sources LISA should be able detect the effects
induced by the spin of the SMBH.  Because the evolution of the inner BBH
depends on the initial conditions, in a future work we will conduct a thorough
survey of the parameter space and identify those triple systems where the
effects of the spin of the SMBHs are detectable by LISA.
\acknowledgments 

This work is supported by the NSFC grants No. 11690021, 11575271, 11747601,
11873022.  QGH acknowledges support by the Strategic Priority Research Program
of the Chinese Academy of Sciences (Grant No. XDB23000000, XDA15020701), and
the Top-Notch Young Talents Program of China.  XC is partly supported by the
Strategic Priority Research Program of the Chinese Academy of Sciences through
the grants No. XDB23040100 and XDB23010200.

\bibliographystyle{apj.bst}

\end{document}